\documentclass[journal]{IEEEtran}

\usepackage{array}
\newcolumntype{L}[1]{>{\raggedright\let\newline\\\arraybackslash\hspace{0pt}}m{#1}}
\newcolumntype{C}[1]{>{\centering\let\newline\\\arraybackslash\hspace{0pt}}m{#1}}
\newcolumntype{R}[1]{>{\raggedleft\let\newline\\\arraybackslash\hspace{0pt}}m{#1}}

\usepackage[ruled,vlined]{algorithm2e}
\usepackage{slashbox}
\usepackage{dblfloatfix}
\usepackage{amsmath,amssymb,amsfonts}
\usepackage{graphicx} 
\usepackage{textcomp}
\usepackage{array}
\usepackage{pifont}
\usepackage{footnote}
\usepackage{xr-hyper}
\usepackage{hyperref}

\newcolumntype{P}[1]{>{\centering\arraybackslash}p{#1}}

\usepackage[table, dvipsnames]{xcolor}

\makeatletter

\begin{document}
\title {Golden Reference-Free Hardware Trojan Localization using Graph Convolutional Network}
\author{Rozhin~Yasaei,~\IEEEmembership{Graduate Student Member,~IEEE,}
        Sina~Faezi,~\IEEEmembership{Graduate Student,~IEEE,}
        and~Mohammad~Abdullah~Al Faruque,~\IEEEmembership{Senior Member,~IEEE}}
% \thanks{M. Shell was with the Department
% of Electrical and Computer Engineering, Georgia Institute of Technology, Atlanta,
% GA, 30332 USA e-mail: (see http://www.michaelshell.org/contact.html).}% <-this % stops a space
% \thanks{J. Doe and J. Doe are with Anonymous University.}% <-this % stops a space
% \thanks{Manuscript received April 19, 2005; revised August 26, 2015.}}

\markboth{IEEE TRANSACTIONS ON Very Large Scale Integration Systems}%
{Yasaei \MakeLowercase{\textit{et al.}}: Golden Reference-Free Hardware Trojan Localization using Graph Convolutional Network}
\maketitle 

\begin{abstract}
The globalization of the Integrated Circuit (IC) supply chain has moved most of the design, fabrication, and testing process from a single trusted entity to various untrusted third-party entities worldwide. The risk of using untrusted third-Party Intellectual Property (3PIP) is the possibility for adversaries to insert malicious modifications known as Hardware Trojans (HTs). These HTs can compromise the integrity, deteriorate the performance, deny the service, and alter the functionality of the design. While numerous HT detection methods have been proposed in the literature, the crucial task of HT localization is overlooked. Moreover, a few existing HT localization methods have several weaknesses: reliance on a golden reference, inability to generalize for all types of HT, lack of scalability, low localization resolution, and manual feature engineering/property definition.
To overcome their shortcomings, we propose a novel, golden reference-free HT localization method at the pre-silicon stage by leveraging Graph Convolutional Network (GCN). In this work, we convert the circuit design to its intrinsic data structure, graph and extract the node attributes. Afterward, the graph convolution performs automatic feature extraction for nodes to classify the nodes as Trojan or benign. 
Our approach is automated and does not burden the designer with manual code review. It locates the Trojan signals with 99.6\% accuracy, 93.1\% F1-score, and a false-positive rate below 0.009\%.
\end{abstract}

%%%%%%%%%%%%%%%%%%%%%%%%%%%%%%%%%%%%%%%%%%%%%%%%%%%%%%%%%%%%%%%%%%%%%%%%%%%%%%%%%

%%%%%%%%%%%%%%%%%%%%%%%%%%%%%%%%%%%%%%%%%%%%%%%%%%%%%%%%%%%%%%%%%%%%%%%%%%%%%%%%%
\vspace{-2em}
\section{Introduction}

\begin{figure*}[t]
\centering
\includegraphics[width=0.96\textwidth]{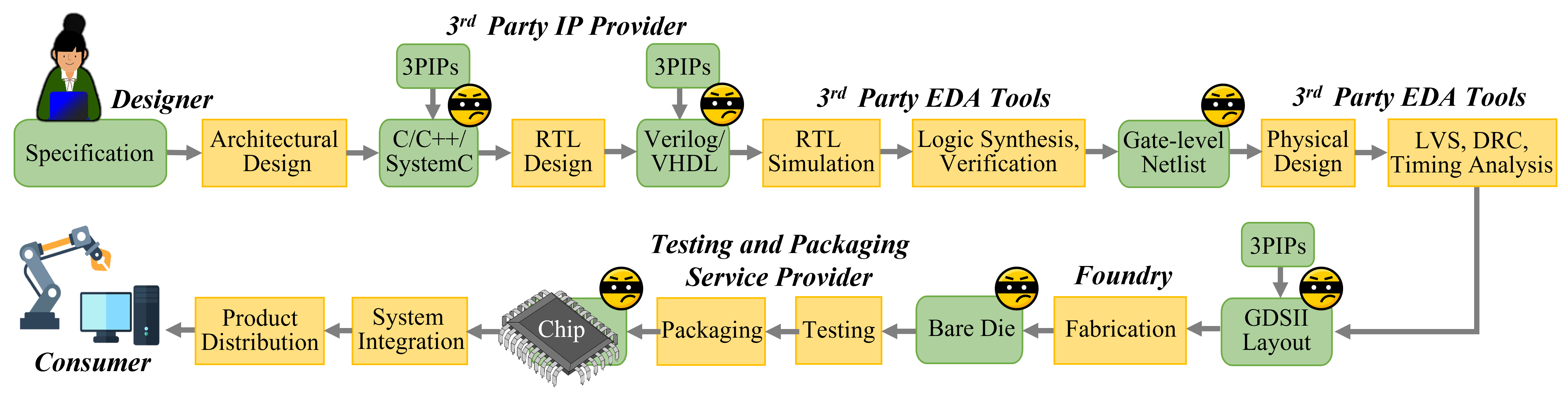}
\vspace{-1.5em}
\caption{IC supply chain which is vulnerable to HT insertion in different stages.}
\vspace{-1.5em}
\label{fig:supply-chain}
\end{figure*}

The growing complexity of Integrated Circuits (IC), time-to-market pressure, and expensive design and manufacturing processes have promoted the globalization of the semiconductor industry. Outsourcing the fabrication and depending on third-party hardware Intellectual Property (IP) blocks and Electronic design automation (EDA) tools raise the risk of intentional and malicious manipulation of the circuit, known as Hardware Trojan (HT). Figure \ref{fig:supply-chain} demonstrates the IC supply chain and the involving parties which are vulnerable points of HT insertion.
Currently, HT is a significant hardware security concern with devastating consequences such as denial of service, malfunctioning, data leakage, and performance degradation in the chip. The attackers usually design HT to be a tiny circuit hidden inside the main design, normally inactive with minimal effect on the chip's functionality and specification. The HT often gets triggered under rare circumstances, and consequently, it can escape detection by routine simulation and functional testing. 

Trojan detection is crucial to ascertain the authenticity of 3PIPs and prevent the negative consequences of HTs. However, HT detection does not suffice to ensure the fabrication of a trustworthy chip, and HT localization is the next essential step. The hardware IPs fall into three classes based on their format and level of abstraction; Soft IP (i.e., synthesizable Verilog or VHDL source code), Firm IP (i.e., placed RTL block and netlist), and  Hard IP (i.e., physical layout and GDSII). Soft IP is the most popular IP core, and IP trust revolves around it. However, it has the most vulnerability against HT insertion because the flexibility and high level of abstraction in Register Transfer Level (RTL) codes facilitate the HT design and implementation for the attacker \cite{tehranipoor2014integrated}.

Manual review of hardware design to pinpoint HT is very time-consuming and error-prone, especially for an industrial-level large design. Due to the paramount importance of the HT threat, numerous defense mechanisms are proposed in the literature to determine if the design is infested with HT. Still, they fail to locate it in the IC design. 
Some works analyze parameters, and side-channel data of circuits such as polynomials of gate-level implementation \cite{farahmandi2017trojan}, thermal map \cite{tang2019activity}, or path delays \cite{sabri2021sat} to pinpoint the disturbance introduced by HT. They have the premise that a trusted Trojan-free reference design called golden reference exists to compare against the parameters of the circuit under test, which is an unrealistic assumption. In order to obtain the golden reference, the whole process of IC design, test, and fabrication should be performed by in-house trusted teams, EDA tools, and manufacturing facilities that would be very expensive and infeasible in practice.
%golden reference is very expensive to fabricate and it is even infeasible to obtain in the pre-silicon stage.

%%%%%%%%%%%%%%%%%%%%%%%%%%%%%%%%%%%%%%%%%%%%%%%%%%%%%%%%%%%%%%%%%%%%%%%%%%%%%%%%%
%%%%%%%%%%%%%%%%%%%%%%%%%%%%%%%%%%%%%%%%%%%%%%%%%%%%%%%%%%%%%%%%%%%%%%%%%%%%%%%%%

HT defense methods based on formal verification and code analysis define some properties for HT, analyze the hardware design, and mark the areas satisfying the predefined properties. For example, \cite{islam2020framework} examines RTL codes using word-level statistics of the inputs and tags the arithmetic blocks with rare nets as vulnerable to HT. \cite{sturton2011defeating} flags the unused portion of the circuit as malicious. \cite{waksman2013fanci} measures the degree of control that an input signal has on the operation and outputs of the circuit and marks weakly-affecting inputs as possible HT triggers.
These works narrow down the search space for HT. However, they still burden the designer to manually review the suspicious areas, which can be a large circuit due to low localization resolution. Moreover, most of the existing solutions require manual property definition or feature extraction, and they fail to outline a comprehensive set of properties or features representing all kinds of HTs. Consequently, they are effective only for particular HTs.

There is an increasing trend to explore the graph representation of hardware for security purposes \cite{fyrbiak2019graph} because hardware design is a non-Eulicidian structural data that shares similar properties with a graph. The graph is a mathematical structure that represents the relation between pairs of objects. It preserves the topological information that makes it the best match for modeling the fundamental objects in the hardware design process \cite{ma2020understanding}. For instance, the graph is leveraged to represent the hardware design in \cite{piccolboni2017efficient, fyrbiak2019graph} for HT mitigation. Still, Graph matching algorithms fail to recognize unknown HTs and are not scalable to large designs due to high complexity.

Deep learning has introduced potent techniques that revolutionized many fields of study \cite{yasaei2020iot}, but it operates on Euclidean data and cannot be directly used for hardware design. The current deep learning models for HT mitigation examine the side-channel emissions of the fabricated chip, which are time-series data \cite{htnet}. To fill the gap and apply machine learning to hardware design, we convert the design from the textual format of HDL code to a graph and leverage GCN, which is like deep learning operating on graphs. A recent work \cite{gnn4tj} proposes a graph classification model based on a graph neural network to find whether the circuit is infested with HT or not, but it fails to locate the Trojan.

In order to overcome the shortcomings of current approaches, we propose a novel, golden reference-free HT localization method in the pre-silicon stage.  We generate a graph representation for the hardware design,  assign attributes to nodes in the graph, classify nodes, and locate the Trojan in HDL code based on the graph node's class. Our node classification model is based on GCN that automatically aggregates the features in graph nodes through the graph convolution operation. 
We create a dataset of hardware designs by inserting HT benchmarks from TrustHub \cite{trusthub} to different circuits. Our methodology is trained on this dataset to learn the behavior and features of Trojan nodes. Then, the trained model can locate the Trojan nodes based on their malicious abnormal behavior in even new and unknown Trojans in a fully automated process without any need for manual review. 

%%%%%%%%%%%%%%%%%%%%%%%%%%%%%%%%%%%%%%%%%%%%%%%%%%%%%%%%%%%%%%%%%%%%%%%%%%%%%%%%%
\subsection{Research Challenges}

HT detection and localization is a difficult problem and the current solutions suffer from the following shortcomings:
\begin{itemize}

\item  \textbf{Reliance on golden reference}: A Trojan-free circuit called golden reference for comparison with the circuit under test is not available in the real world, and the golden reference-dependent methods are not practical.

\item  \textbf{Unable to generalize}: Various types of HTs are discovered so far, and new HT designs are continuously introduced. Due to the variety in HT design and specification, defining a template or some properties that describe all HTs is challenging. Consequently, many countermeasures fail to generalize and are limited to known HTs or only HTs with a specific trigger or payload.

\item  \textbf{Low localization resolution}: Some works output the areas of the circuit that are vulnerable to HT insertion and due to low localization resolution, they burden the designer with an exhaustive review of suspicious regions.

\item  \textbf{Manual feature extraction}: Algorithmic and classic machine learning approaches rely on an expert to define properties and extract features from the circuit, which is error-prone and exhausting. 

\item \textbf{Scalability}: With the increased complexity of ICs, scalability has become an essential characteristic of any circuit analysis tool, but complex algorithms fail to scale for large designs.

\end{itemize}
%%%%%%%%%%%%%%%%%%%%%%%%%%%%%%%%%%%%%%%%%%%%%%%%%%%%%%%%%%%%%%%%%%%%%%%%%%%%%%%%%
\vspace{-1.5em}
\subsection{Contributions}

In this paper, we surmount the aforementioned research challenges and propose a novel, golden reference-free approach for HT localization that is fully automated with no need for manual revision by experts. To the best of our knowledge, this is the first work to apply GCN for HT localization. Our contributions can be summarized as:

\begin{itemize}
    \item The hardware design HDL code is converted to data-flow graph using hardware design toolkit \cite{pyverilog}. We develop an algorithm to extract the attributes of nodes and assign an attribute vector to each one.
    
    \item We construct a node classification model based on GCN that automatically aggregates the features for each node in the graph representation of hardware design, learns their behavior, and marks the malicious nodes.  
    
    \item We develop a Trojan labeling algorithm that provides a mapping from HDL code to its graph and labels the nodes in the graph as Trojan or benign. This algorithm determines the HT label vector of the training dataset, which is deployed by the GCN model for training and calculating classification loss.
    
    \item We survey the existing pre-silicon HT detection and localization methods and their shortcomings to picture the current state and challenges of this research area as well as the potential of graph learning for advancement in hardware security.
    
\end{itemize}

%%%%%%%%%%%%%%%%%%%%%%%%%%%%%%%%%%%%%%%%%%%%%%%%%%%%%%%%%%%%%%%%%%%%%%%%%%%%%%%%%
\section{Related Works and Backgrounds} 
\label{sec:related-works} 

%%%%%%%%%%%%%%%%%%%%%%%%%%%%%%%%%%%%%%%%%%%%%%%%%%%%%%%%%%%%%%%%%%%%%%%%%%%%%%%%%
\subsection{Hardware Trojan Detection}

Due to the severity of the HT threat, its detection is extensively studied in the literature, while the importance of HT localization is overlooked. In this section, we survey the approaches that answer whether the design is contaminated with HT but cannot locate the Trojan.

\subsubsection{\textbf{Graph similarity algorithm}}: Graph is shown to be a natural and potent representation of hardware design \cite{ma2020understanding}. In this regard, \cite{piccolboni2017efficient} attempt to detect HT by examination of the similarity between data/control flow graphs of the Trojan and the circuit under test.  \cite{fyrbiak2019graph} further advances the graph matching algorithm by creating a new graph similarity heuristic for hardware security applications. In addition to the lack of scalability and high complexity of graph matching algorithm (NP-complete), this approach is challenged with new types of HT as their detection range is limited to known HTs and fails to generalize. 

\subsubsection{\textbf{Machine learning}}: Machine learning introduces powerful data-driven models that, through the iterative process of training on a dataset, perform optimization to minimize the prediction error and learn to infer correct prediction for a test subject. Since most machine learning models work with Euclidean data, they are mainly used in the side-channel analysis for post-silicon hardware security applications \cite{htm, htnet, ashraf2018towards}. Machine learning models for HT detection in pre-silicon are proposed that are trained on the features extracted from the circuit or its graph representation. For instance, Trojan-net features derived from gate-level netlist are exploited by support vector machine \cite{hasegawa2016hardware} and multi-layer neural network \cite{hasegawa2017hardware}.  Others apply gradient boosting \cite{han2019hardware}, probabilistic neural network \cite{demrozi2017exploiting},  or high-level behavior classification model of artificial immune system \cite{zareen2018detecting} on the features extracted from abstract syntax tree, control flow graph, and data/control-flow graphs of RTL design, respectively. Reliance on a golden reference is one of the shortcomings of many classification models.
Moreover, another common problem with classic machine learning models is the manual feature extraction which is error-prone and highly depends on the developer's knowledge of hardware design. Deep learning models eliminate this problem with automated feature extraction, which is recently employed by \cite{gnn4tj}. It classifies the data-flow graph of RTL design using a graph neural network.  

\subsubsection{\textbf{Formal verification}}: Formal verification translates the circuit to proof checking format and proves it satisfies the pre-defined security properties. The limited detection range is the main weakness of this method as it is difficult to define a set of properties that generalizes to the various types of HT. For example, \cite{rajendran2015detecting} and even its advanced version \cite{rajendran2016formal} can only expose HT payloads that leak data. 
Information flow tracking is a formal method for security verification, deployed in \cite{ardeshiricham2017register, nahiyan2017hardware, hu2018property}. The model checking used in formal verification approaches encounters state explosion and cannot scale for large designs. 

\subsubsection{\textbf{Test pattern generation}}: HT evades detection by the routine verification test and simulations because it is a tiny circuit with negligible effect on circuit functionality and specification. Thus, test pattern generation methods produce a set of test vectors more likely to activate the HT \cite{saha2015improved, zhao2018hardware}. The stand-alone method is usually insufficient, and it is bundled with other techniques such as side-channel analysis \cite{lyu2020automated}.

%%%%%%%%%%%%%%%%%%%%%%%%%%%%%%%%%%%%%%%%%%%%%%%%%%%%%%%%%%%%%%%%%%%%%%%%%%%%%%%%%
\subsection{Hardware Trojan Localization}

The majority of defense mechanisms against HT focus on its detection, and there are inadequate works with the capability to locate the Trojan circuit. For example, \cite{sabri2021sat, tang2019activity} perform HT localization with the assumptions that the design pipeline is trusted, and the attacker resides in the foundry. Both works have the unrealistic premise that a golden reference is available.
\cite{sabri2021sat} proposes an SAT-based test pattern generation scheme that detects and locates the Trojan inserted by foundry by comparing the timing and path delays of the suspicious IC with a golden IC.  
\cite{tang2019activity} extracts Trojan activity factor from the redundant thermal map and performs HT localization by comparing the thermal side-channel of the target chip with the golden reference.

Code analysis is one of the conventional pre-silicon HT defense mechanisms that inspects the HDL code to ascertain suspicious signals in the circuit, and it is mainly restricted to combinational logic. In this technique, the code is scanned based on coverage metrics (toggle, line, state, etc.) to find the potential areas of HT presence. Different methods propose various definitions for a suspicious area, such as unused circuit identification \cite{hicks2010overcoming}, weakly affecting inputs \cite{waksman2013fanci}, and low dependence on functional inputs \cite{zhang2015veritrust}. Due to exclusive definition of HT, later \cite{sturton2011defeating} defeats \cite{hicks2010overcoming} and  \cite{zhang2015veritrust} and \cite{waksman2013fanci} get bypassed by the new HT attack \cite{zhang2014detrust}. \cite{islam2020framework} proposes a framework to analyze RTL codes using word-level statistics of the inputs. It locates the arithmetic blocks with rare nets to be reviewed as candidates vulnerable to HT and can only identify HTs that are always on or triggered by current inputs. Code analysis suggests the circuit areas susceptible to HT insertion and cannot actually locate the HT or even guarantee its detection. 

\cite{farahmandi2017trojan} introduces a formal method based on symbolic algebra by extracting polynomials from the gate-level implementation of the untrustworthy IP and comparing them with the golden reference polynomials.
\cite{islam2019socio} leverages principal features of social network analysis to outline the relation between design properties and locate HT. This approach applies only to combinational Trojans.
%%%%%%%%%%%%%%%%%%%%%%%%%%%%%%%%%%%%%%%%%%%%%%%%%%%%%%%%%%%%%%%%%%%%%%%%%%%%%%%%%
\begin{figure*}[t]
\centering
\includegraphics[width=0.96\textwidth]{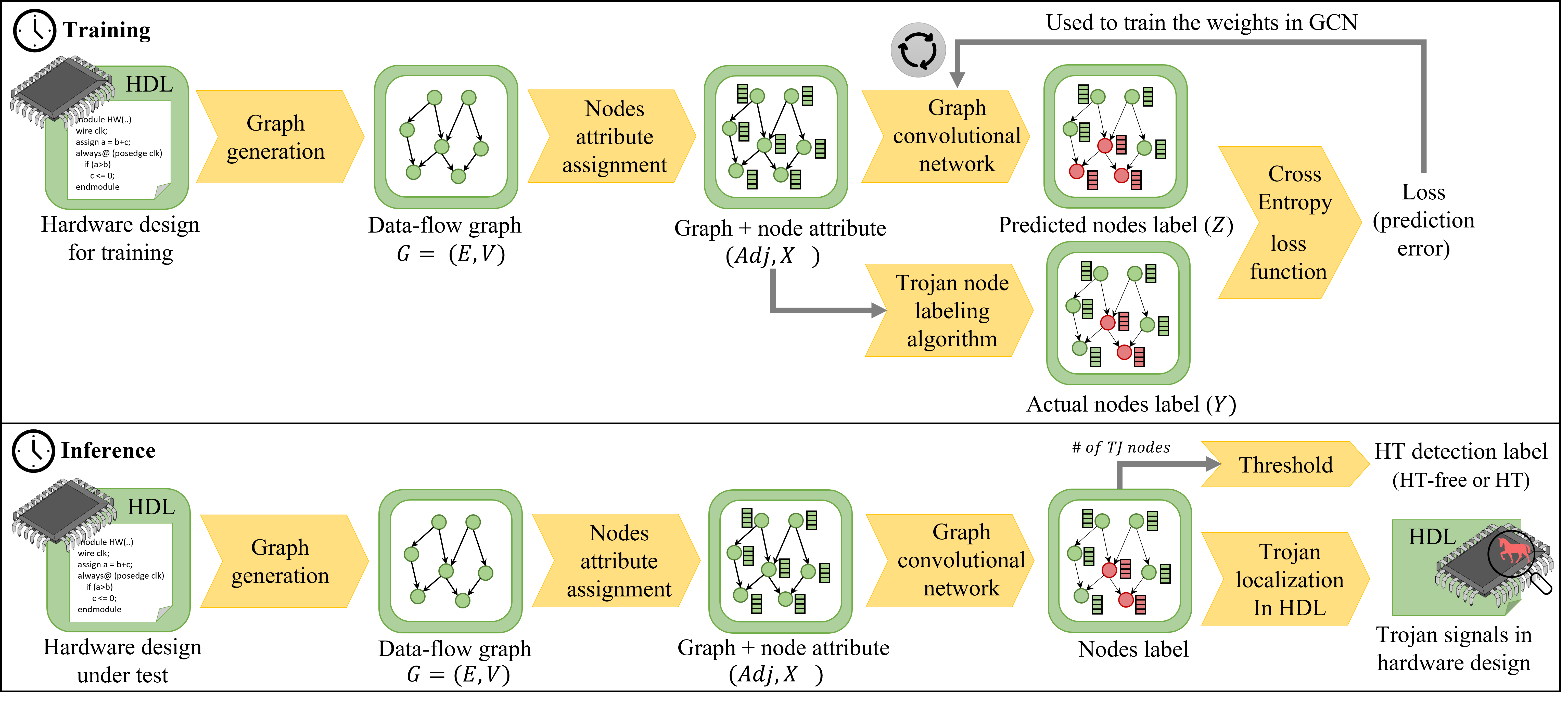}
\vspace{-1em}
\caption{Overview of our HT localization methodology in training and inference phases.}
\vspace{-1.5em}
\label{fig:overview}
\end{figure*}
%%%%%%%%%%%%%%%%%%%%%%%%%%%%%%%%%%%%%%%%%%%%%%%%%%%%%%%%%%%%%%%%%%%%%%%%%%%%%%%%%
%%%%%%%%%%%%%%%%%%%%%%%%%%%%%%%%%%%%%%%%%%%%%%%%%%%%%%%%%%%%%%%%%%%%%%%%%%%%%%%%%
\subsection{Graph in Hardware Applications}

Hardware design is non-Eulicidian structural data that shares similar properties with a graph. The graph is a mathematical structure that represents the relation between pairs of objects. It preserves the topological information that makes it the best match for modeling the fundamental objects in the hardware design process. Thus, the graph is leveraged to represent the hardware in numerous Electronic Design Automation (EDA) problems which shift the problem to choosing the appropriate algorithm from the many well-known graph algorithms and apply it directly or with a slight change to solve the problem. However, developing an effective approach for each problem is still challenging.
Furthermore, many problems are NP-hard with large sizes which makes efficiency a major concern and leads to scalability issue. To tackle the complexity issue, data-driven learning techniques have grabbed much attention. The classical machine learning models include an initial step of manual feature extraction which is followed by model training based on a large set of data instances \cite{ma2020understanding}.

The next generation of machine learning models leverages convolutional layers in deep learning models, making the feature extraction process automated through learning. Recently, deep learning models are developed with high resiliency against adversarial attacks \cite{ashraf2020r2ad}. Although that deep learning has improved the performance in various applications, it cannot be directly applied to graphs because it is originally developed for Euclidean data, and notable extra endeavors are needed to extract features from graphs and encode the structural information. In response, the graph learning method is introduced, which defines the convolutional operation on graphs and automates the feature extraction from graphs. There have been a few works investigating the advantages of graph learning for the hardware security \cite{gnn4tj, gnn4ip, hw2vec} and hardware design automation such as test point insertion \cite{ma2019high}, and power estimation in simulation \cite{zhang2020grannite}. In this work, we leverage a state-of-the-art machine learning model, GCN, to model the hardware for security purposes.
%%%%%%%%%%%%%%%%%%%%%%%%%%%%%%%%%%%%%%%%%%%%%%%%%%%%%%%%%%%%%%%%%%%%%%%%%%%%%%%%%
\section{Methodology}

In this paper, we propose an automated pipeline to locate Trojan circuit at RTL that includes several steps, as depicted in Figure~\ref{fig:overview}; i) converting hardware design to graph, ii) extracting the node attributes, iii) labeling Trojan nodes, iv) node classification, and v) HT localization in HDL. In the following sections, we define our problem formulation and threat model and then, we elaborate on the aforementioned steps for localization.
%%%%%%%%%%%%%%%%%%%%%%%%%%%%%%%%%%%%%%%%%%%%%%%%%%%%%%%%%%%%%%%%%%%%%%%%%%%%%%%%%
\subsection{Problem Formulation and Threat Model}

The main target of our methodology is to locate the Trojan circuit inside hardware design at RTL. The model's input is an HDL code which is later converted to a graph. The graph representation is further processed, and the graph learning model classifies the graph nodes as Trojan or benign. Eventually, the model outputs a list of malicious signals and operations in the HDL code corresponding to Trojan nodes in the graph. 

The graph learning model, GCN, is trained on a dataset of graphs derived from HT benchmarks in which the labels of the nodes are known. Our dataset only includes the HT-infested designs, not any Trojan-free design. Our approach is golden reference-free and able to perform HT localization on unknown HTs. To demonstrate these characteristics, we train our model on a set of circuits and test it on the circuits not observed by the model before in the training stage. Therefore, the model locates Trojan nodes in the circuit under test while it has not seen its golden reference or HT benchmark. Moreover, we make no assumption about the HT payload or trigger type, and the fundamental features of Trojan nodes are automatically aggregated and learned by convolutional layers in our GCN.

An attacker may manipulate the hardware design at any pre-silicon stage of the IC supply chain in our threat model (refer to Figure \ref{fig:supply-chain}), but eventually, the HDL code should be available for our methodology to perform HT localization. Therefore, multiple attack scenarios are feasible. The attacker can be a rogue in-house designer, an untrusted 3PIP design company, or a 3P-EDA tool provider who tampers with the HDL code. The adversary may alter the design in the low level of abstractions, such as netlist and physical layout.  In this case, we assume that the RTL code is obtained by reverse engineering.
%%%%%%%%%%%%%%%%%%%%%%%%%%%%%%%%%%%%%%%%%%%%%%%%%%%%%%%%%%%%%%%%%%%%%%%%%%%%%%%%%
\subsection{Hardware Design Conversion to Graph}

A circuit is described using Hardware Description Languages (HDL) at the design stage, such as Verilog and VHDL. The HDL code has a textual format with predetermined syntax and cannot be directly used as data for machine learning. Thus, we convert the HDL code to a graph that embeds the design features and preserves the topological information. 

\begin{figure}[t]
\centering
\includegraphics[width=0.46\textwidth]{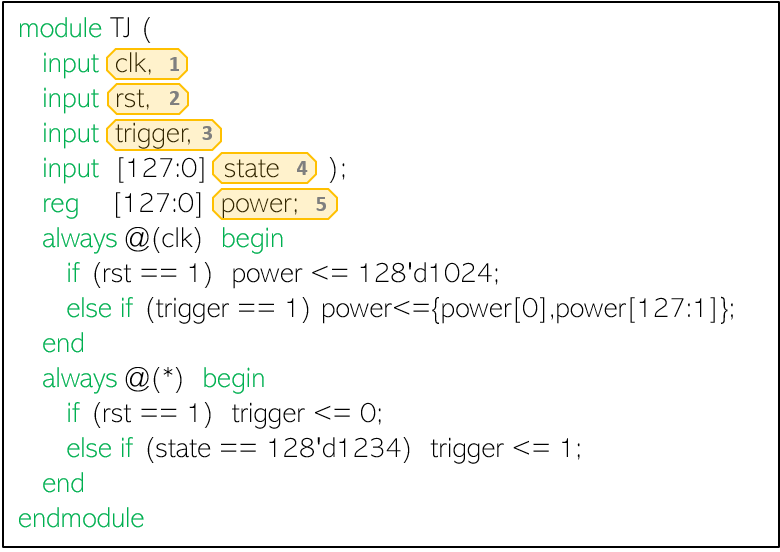}
\vspace{-1em}
\caption{The Verilog code of AES-T1800 Trojan benchmark.}
\vspace{-1.5em}
\label{fig:TJcode}
\end{figure}

\begin{figure}[t]
\centering
\includegraphics[width=0.46\textwidth]{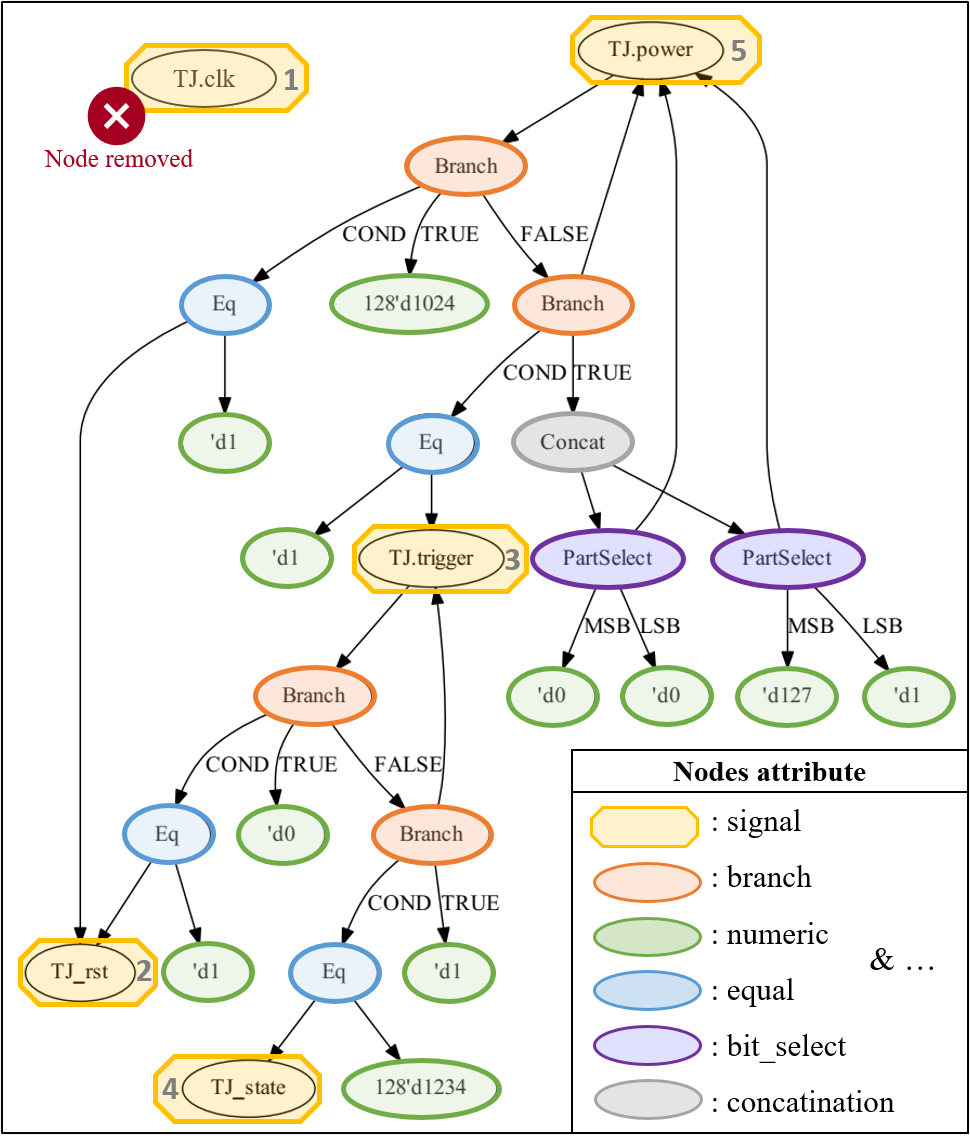}
\vspace{-1em}
\caption{The data-flow graph and node attributes of AES-T1800 Trojan benchmark, shown in Figure~\ref{fig:TJcode}.}
\vspace{-1.5em}
\label{fig:TJgraph}
 
\end{figure}

HDL code comprises modules, signals, and operations. Modules are used to cluster parts of circuits and better express the hierarchy in the hardware design, but they do not affect the design specification. On the other hand, signals and operations fundamentally describe the hardware design. A  signal can be a register or wire in HDL code, and it carries a value that is changed through an operation or assignment. For instance, the Verilog code for the AES-T1800 HT benchmark is shown in Figure~\ref{fig:TJcode} with its corresponding data-flow graph in Figure~\ref{fig:TJgraph}.
 
To convert HDL to graph, we combine the modules to have a single HDL code for the whole design. Afterward, we parse it and extract the data dependency subgraphs for all signals in HDL code using a hardware design tool called PyVerilog \cite{pyverilog}. Each signal subgraph is a tree that expresses how the value of root signals depends on the operations and other signals in the design. We connect the common nodes in the subgraphs to construct one data-flow graph per hardware. The extracted graph $G = (V, E)$ is a directed homogeneous graph in which each node is named after its corresponding signal/operation in the HDL code, and an edge $e_{ij}$ indicates that node $v_i$ depends on node $v_j$. Lastly, the graph is processed to remove standalone nodes such as \textit{clk} node in the data-flow graph of AES-T1800 HT benchmark, depicted in Figure~\ref{fig:TJgraph}.
%%%%%%%%%%%%%%%%%%%%%%%%%%%%%%%%%%%%%%%%%%%%%%%%%%%%%%%%%%%%%%%%%%%%%%%%%%%%%%%%%
%%%%%%%%%%%%%%%%%%%%%%%%%%%%%%%%%%%%%%%%%%%%%%%%%%%%%%%%%%%%%%%%%%%%%%%%%%%%%%%%%
%%%%%%%%%%%%%%%%%%%%%%%%%%%%%%%%%%%%%%%%%%%%%%%%%%%%%%%%%%%%%%%%%%%%%%%%%%%%%%%%%
%%%%%%%%%%%%%%%%%%%%%%%%%%%%%%%%%%%%%%%%%%%%%%%%%%%%%%%%%%%%%%%%%%%%%%%%%%%%%%%%%
%%%%%%%%%%%%%%%%%%%%%%%%%%%%%%%%%%%%%%%%%%%%%%%%%%%%%%%%%%%%%%%%%%%%%%%%%%%%%%%%%
\begin{figure*}[t]
\centering
\includegraphics[width=0.90\textwidth]{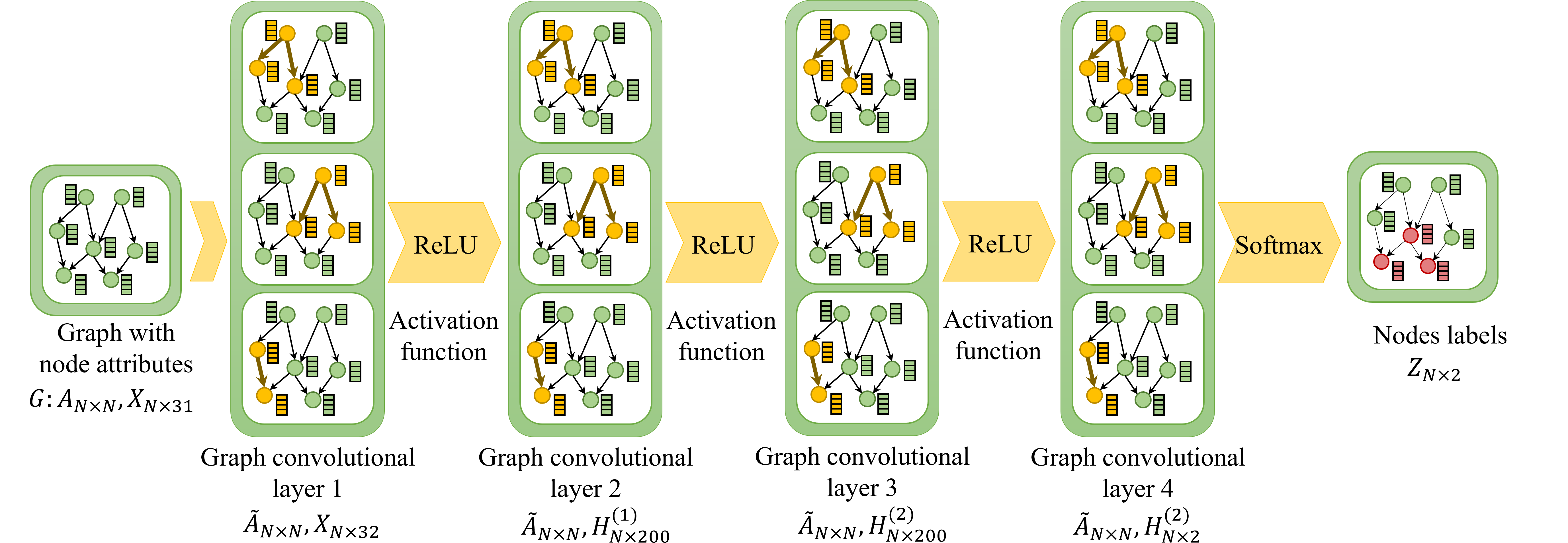}
\vspace{-1em}
\caption{The architecture of our GCN model for node classification.}
\vspace{-1.5em}
\label{fig:gcn}
 
\end{figure*}
%%%%%%%%%%%%%%%%%%%%%%%%%%%%%%%%%%%%%%%%%%%%%%%%%%%%%%%%%%%%%%%%%%%%%%%%%%%%%%%%%
%%%%%%%%%%%%%%%%%%%%%%%%%%%%%%%%%%%%%%%%%%%%%%%%%%%%%%%%%%%%%%%%%%%%%%%%%%%%%%%%%
\subsection{Node Attribute Extraction}

The initial data-flow graph $G = (V, E)$ expresses the flow of information and connections between components in the circuit, but it does not differentiate between the nodes. Therefore, we develop a node analyzer to extract the type of nodes from their name, which Pyverilog generates during graph generation. Then, the analyzer assigns an attribute vector to each node which is further used as an input feature to GCN. Nodes can be categorized as \textit{constant}, \textit{signal}, or \textit{operation}. The \textit{constant} nodes represent numbers and are tagged as numeric regardless of their value. The \textit{signal} nodes are derived from a wire or register in the HDL code and tagged as input, output, or signal based on their position in the circuit. The \textit{operation} nodes are related to the operands and conditional statements in the HDL code. They have a wide variety, including gates (not, and, or, xor, etc.), branch, conditional operands (equal, less than, greater than, etc.), Part select, and concatenation. We have detected 28 different types of \textit{operation} nodes which sums up the total numbers of the node types to 32. The tags are independent of the circuit design and they represent all the possible types of nodes that can be generated by our graph generation pipeline for any HDL code. Some examples of node tags are demonstrated in Figure~\ref{fig:TJgraph}. After tagging nodes, we generate an attribute vector for each node by performing one-hot encoding on tags. Therefore, the new directed graph with $N$ nodes and $F$ different tags is defined by $A \in R^{N \times N}$ and $X \in R^{N \times F}$ where $A$ is an asymmetric adjacency matrix, and $X$ is the matrix of node attributes.
%%%%%%%%%%%%%%%%%%%%%%%%%%%%%%%%%%%%%%%%%%%%%%%%%%%%%%%%%%%%%%%%%%%%%%%%%%%%%%%%%
\subsection{Trojan Labeling Algorithm}

\begin{figure}[t]
\centering
\includegraphics[width=0.46\textwidth]{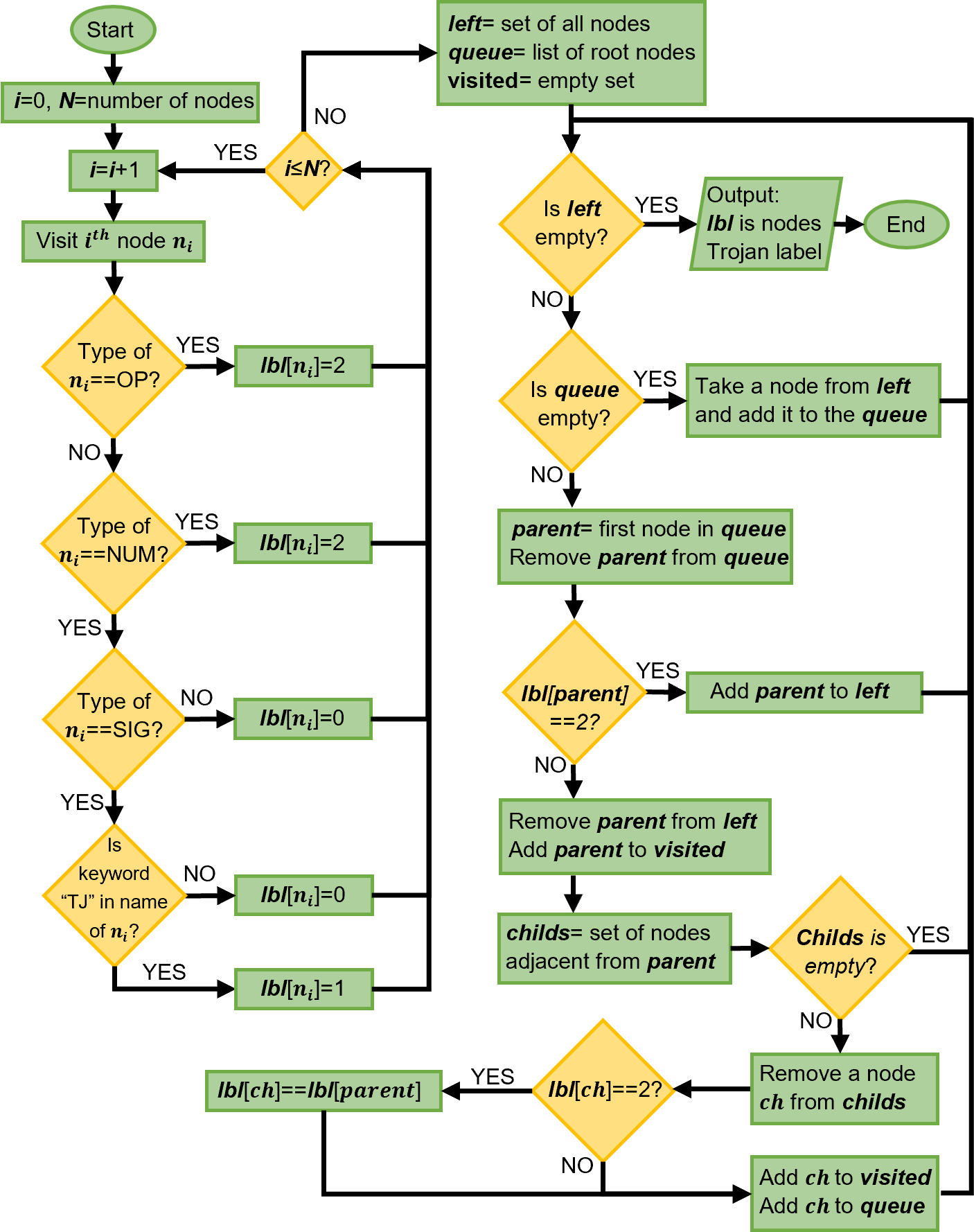}
\vspace{-1em}
\caption{ The flowchart of Trojan labeling algorithm. }
\vspace{-1.5em}
\label{fig:flowchart}
\end{figure}

Although HT circuit is known in HDL codes used for training the GCN, the graph representation of the circuit does not have any notion of Trojan. Therefore, we develop an algorithm to determine the HT nodes in the graph representation of HT benchmarks. The algorithm has an HDL processing step in which a keyword is added to the signals and modules of the HT circuit. This keyword will be visible in the name of \textit{signal} nodes of the HT circuit, but the \textit{constant} and \textit{operation} nodes will not be affected. So, after graph generation, the labeling algorithm iterates among the nodes and flags the \textit{operation} and \textit{constant} nodes as 2 (can be Trojan), the \textit{signal} node with the keyword as 1 (definitely Trojan), and the rest of the \textit{signal} nodes as 0 (not Trojan). Thus, the flag of \textit{signal} nodes is known to be Trojan, or benign. The flag of \textit{constant} and \textit{operation} nodes are modified based on the rules that the \textit{operation} nodes applied to Trojan nodes are part of the HT circuit and the \textit{constant} nodes inherit the flag of their parent \textit{operation} node.
Algorithm~\ref{alg:TJlabeling} and Figure~\ref{fig:flowchart} show how the algorithm traverses the graph starting from root nodes and modifies the number 2 flags based on these rules until there is no flag of 2 left and all nodes are marked either as Trojan or benign. The algorithm results in a Trojan label vector $Y \in {[0, 1]}^N $ for each graph with $N$ nodes in which the malicious nodes are marked as 1. This label vector is further used as the classification label of the training dataset to train the GCN model. Note that this step is only once performed for the training dataset in the training stage, and it is not required in the inference stage when the model is already trained and ready to locate HT in a circuit. 
%%%%%%%%%%%%%%%%%%%%%%%%%%%%%%%%%%%%%%%%%%%%%%%%%%%%%%%%%%%%%%%%%%%%%%%%%%%%%%%%%
\begin{algorithm}[h]
\normalsize
\SetAlgoLined
\SetAlgoVlined
\DontPrintSemicolon
\textbf{Input:} Nodes name list $name$, Nodes type list $type$. \\
\textbf{Input:} Signal nodes tag $SIG$, Operation nodes tag $OP$, Numeric node tag $NUM$. \\
\textbf{Output:} Trojan label vector $lbl$. \\
Initialize $queue$ = a list of root node. \\
Initialize $left$ = a set of all nodes. \\
Initialize $visited$ = an empty set. \\
\ForEach{node $n_i$ in $left$ } {
    \If{ ($type$[$n_i$] is $OP$ or $NUM$)} { 
        $lbl$[$n_i$] = 2  \\
    } \ElseIf { ($type$[$n_i$] is $SIG$) and (TJ in $name$[$n_i$]) } { 
        $lbl$[$n_i$] = 1  \\
    } \Else {
        $lbl$[$n_i$] = 0   \\
    }
}
\While{ $left$ is not empty}{  
    \If {$queue$ is empty } {  
        $node$ = $left$.pop()  \\
        add $node$ to $queue$  \\
    } \Else { 
        $parent$ = $queue$.pop() \\
        \If {$lbl$[$parent$] != 2 } {
            remove $parent$ from $left$  \\
            add $parent$ to $visited$   \\
            \ForEach{ child $c_i$ of $parent$} {   
                \If {$c_i$ not in $visited$}{ 
                    \If {$lbl$[$c_i$] == 2} { 
                        $lbl$[$c_i$] = $lbl$[$parent$] \\
                    }
                    add $c_i$ to $visited$ \\
                    add $c_i$ to $queue$ \\
                }
            }
        }
        \Else { 
            add $parent$ to $left$  \\
        }
    }
}
\Return Trojan label vector $lbl$ 
\caption{Trojan Nodes Labeling Algorithm}
\label{alg:TJlabeling}
\end{algorithm}
%%%%%%%%%%%%%%%%%%%%%%%%%%%%%%%%%%%%%%%%%%%%%%%%%%%%%%%%%%%%%%%%%%%%%%%%%%%%%%%%%

%%%%%%%%%%%%%%%%%%%%%%%%%%%%%%%%%%%%%%%%%%%%%%%%%%%%%%%%%%%%%%%%%%%%%%%%%%%%%%%%%
\subsection{Graph Convolutional Networks}

%The graph representation embeds data of nodes (nodes attribute) and structural information (the connection between nodes). 
% Inspired by deep learning, GCN are devised to embed nodes with different features while taking the topological information into account. Our node classification model is adapted from \cite{kipf2016semi} since the GCN is shown to be a compelling neural network architecture for graph learning in various applications \cite{gao2021gdroid}. 
% 
Traditionally, deep learning models often use an array/stack of trainable filters such as convolutional neural networks to extract meaningful features for grid-like structured data. Inspired by those works, we adopt the GCN layer as our trainable filter from \cite{kipf2016semi}.  GCN is devised to embed nodes with different features while taking the topological information in non-euclidean data into account. The input of the GCN model is a graph $G = (V,E)$ is represented by adjacency matrix $A \in N \times N$ and node attribute matrix $X \in N \times F$ where $N$ is the number of nodes. $F$ is the length of each node attribute vector in our model. 
Each graph convolution layer aggregates information from immediate node neighbors and update nodes through a process called \textit{message passing} based on the following formula:
\begin{equation}
    {H^{(l+1)}} = \sigma(\Tilde{A} {H^{(l)}} W^{(l)}) 
\end{equation}
Here, $l$ denotes the layer number, and $H^{(0)}$ is the initial node features that equal to $X$, the node attributes matrix. $W^{(l)}$ is a layer-specific trainable weight matrix. $\sigma(.)$ denotes the activation function that is Rectified Linear Unit (ReLU) in our model. 
To perform graph convolution, the normalized adjacency matrix $\Tilde{A}$ is computed by:
\begin{equation}
     \Tilde{A} = \widehat{D}^{-\frac{1}{2}} \widehat{A} \widehat{D}^{-\frac{1}{2}}
\vspace{-0.4em}
\end{equation}
where $\widehat{D}$ is the diagonal degree matrix to solve the problem of scale change of the feature vectors after multiplication by the matrix $A$. It is calculated by $\widehat{D} = \sum_{j}\widehat{A}_{ij}$ and $\widehat{A}$ is derived from $\widehat{A} = A + I_N$ where $I_N$ is the identity matrix that adds self-loop connection to $A$, adjacency matrix of graph $G$, to make sure each node embeds its previous value from last iteration as well as new data from its neighbors.

Stacking the graph convolution layers, we create a GCN that is able to integrate information from a larger set of neighbors. Our model architecture is illustrated in Figure~\ref{fig:gcn}. It comprises three convolution layers with a ReLU activation function and one last convolution layer connected to a layer of Softmax units to classify each node as Trojan or benign and generate the predicted node label $Y$. It concludes the computations of our GCN model as below:
\begin{equation}
    Z = Softmax(\Tilde{A} \sigma(\Tilde{A} \sigma(\Tilde{A} \sigma(\Tilde{A} X W^{(0)}) W^{(1)}) W^{(2)} ) W^{(3)} )
\end{equation}
where $Z \in {[0,1]}^2$ indicates the predicted node labels in which $[1,0]$ denotes Trojan and $[0,1]$ denotes benign node.

%The last graph convolution layer is added to decrease the number of features for each node to the number class (which is 2 in this case), and instead of the ReLU activation function, it is followed by a layer of Softmax units to classify each node as Trojan or benign and generate the predicted node label $Y$. It concludes the computations of our GCN model as below:

\textbf{Training}: Figure~\ref{fig:overview} summarize the training process and units. The first step in training is preparing a training dataset based on the GCN model requirement. Thus, the HDL codes are converted to data-flow graphs in which the nodes have been assigned an attribute vector and a label (Trojan/benign). The attribute vectors feed information about each node's characteristic to the model, and the labels are used to calculate the error due to the node misclassification.
Training is an iterative optimization process that modifies the weights in GCN to minimize its classification error, anointed as the loss. We use Adam optimizer \cite{kingma2014adam}, a conventional optimization technique for efficient gradient descent to minimize the loss $L$. We utilize the cross-entropy loss function to calculate the error over all  nodes in a graph using this formula:
\vspace{-0.4em}
\begin{equation}
    L = - \sum_{i \in V} \sum_{j=0}^{C} Y_{ij} ln(Z_{ij})
\end{equation}
\vspace{-0.2em}
where $C$ is the number of classes which is two (Trojan and benign), and the $j$ indicated the dimension of the output vector. $V$ is the set of nodes in a graph, and $i$ iterates over them. $Y$ is the actual label of nodes obtained from the Trojan node labeling algorithm, and $Z$ is the predicted label. Note that $L$ is node classification loss for one graph, and total loss is the summation of all graphs loss.

\textbf{Inference}: At the inference stage that the model is trained, we use it to test new hardware design, as shown in Figure~\ref{fig:overview}. After label prediction, the node labels are passed to HT localization in the HDL unit. It employs the mapping between graph nodes and HDL signals to mark malicious signals in HDL code based on Trojan nodes in the graph. We also perform HT detection by counting the number of nodes predicted to be Trojan and label the circuit as Trojan-free if the number of Trojan nodes is lower than a threshold and, basically, negligible compared to the size of the design. The user can set the threshold depending on their target sensitivity. 

%%%%%%%%%%%%%%%%%%%%%%%%%%%%%%%%%%%%%%%%%%%%%%%%%%%%%%%%%%%%%%%%%%%%%%%%%%%%%%%%%
\begin{table*}[]
\caption{ Comparing the HT localization methods in the literature. }
\vspace{-1em}
\label{tab:survey}
\begin{tabular}{|l|c|c|c|c|c|l|}
\hline
\rowcolor[HTML]{AED09D} 
\multicolumn{1}{|c|}{\cellcolor[HTML]{AED09D}Method} & Stage & \begin{tabular}[c]{@{}c@{}}Golden\\ chip-Free\end{tabular} & Automated & \begin{tabular}[c]{@{}c@{}}Localization\\ Resolution\end{tabular} & \begin{tabular}[c]{@{}c@{}}HT\\ diversity\end{tabular} & \multicolumn{1}{c|}{\cellcolor[HTML]{AED09D}Performance} \\ \hline
\rowcolor[HTML]{EBF1E9} 
GCN (ours) & Pre-S$^2$ & Yes & Yes & High & High &  HT localization with 99.6\% accuracy and 98.9\% precision  \\ \hline
Social network \cite{islam2019socio} & Pre-S & Yes & No & High & Low &  HT localization with 97.3\% accuracy and less than 2\% false positive  \\ \hline
\rowcolor[HTML]{EBF1E9} 
Code analysis \cite{islam2020framework} & Pre-S & Yes & No & Low & Low &  Activity estimation with less than 2\% error to flag low-activity as HT  \\ \hline
VeriTrust \cite{zhang2015veritrust} & Pre-S & Yes & No & Low & Low &  HT localization with 100\% recall and 11.5\% precision  \\ \hline
\rowcolor[HTML]{EBF1E9} 
FANCI \cite{waksman2013fanci} & Pre-S & Yes & No & Low & Low &  HT localization with 100\% recall and less than 8\% false positive  \\ \hline
UCI \cite{hicks2010overcoming} & Pre-S & Yes & No & Low & Low &  HT localization with 100\% recall and 7.5\% precision  \\ \hline
\rowcolor[HTML]{EBF1E9} 
Symbolic algebra\cite{farahmandi2017trojan} & Pre-S & No & No & High & High &  HT localization with 100\% recall and 74\% precision  \\ \hline
Thermal map \cite{tang2019activity} & Post-S$^3$ & No & No & Low & High &  Successfully locates the HTs with less than 20 gates  \\ \hline
\rowcolor[HTML]{EBF1E9} 
Path delay \cite{sabri2021sat} & Post-S & No & No & High & High &  HT localization with 100\% recall and 0.56\% false positive rate  \\ \hline
\end{tabular}
\vspace{-2.5em}
\end{table*}
%%%%%%%%%%%%%%%%%%%%%%%%%%%%%%%%%%%%%%%%%%%%%%%%%%%%%%%%%%%%%%%%%%%%%%%%%%%%%%%%%
%%%%%%%%%%%%%%%%%%%%%%%%%%%%%%%%%%%%%%%%%%%%%%%%%%%%%%%%%%%%%%%%%%%%%%%%%%%%%%%%%
\vspace{-0.5em}
\section{Evaluation}

%%%%%%%%%%%%%%%%%%%%%%%%%%%%%%%%%%%%%%%%%%%%%%%%%%%%%%%%%%%%%%%%%%%%%%%%%%%%%%%%%
\subsection{Experimental Setup}

We construct and assess our GCN model on the graph representation of a dataset, consisted of 49 Trojan-infested RTL codes that are listed in Table~\ref{tab:resultPerHW}. The limited number of graphs in our dataset is not problematic since our machine learning model is for node classification, and each graph contains thousands of nodes, refer to Table~\ref{tab:ave-result}. 
 An extensive dataset enhances the model's performance and capability to learn a generic knowledge of HT and learning-based model is easily adaptable by adding new circuits and Trojans to training for further generalization. 
Our dataset comprises three base circuits that contain various HTs. AES, DES, and RC5 are encryption cores with different algorithms that get an input number as plaintext along with a secret key and output the encrypted number known as ciphertext. The AES samples are derived from the TrustHub benchmark \cite{trusthub} which is the most popular open hardware Trojan datasets used in the literature. The RC5 and DES are open-source designs in which we insert the Trojan circuits extracted from AES-Txx benchmarks. However, some of TrustHub HT benchmarks are specific to AES and cannot be inserted in RC5 or DES circuits due to dependency on the internal signals of AES. In Table~\ref{tab:ave-result}, the first part of benchmark name represents the base circuit and the second part, shows the type of Trojan inserted in the base circuit. For example, DES-T100 shows DES circuit infected with T100 Trojan from TrustHub dataset.  
All the algorithms and models are implemented in the Python language. We use PyTorch and the Geometric extension library to build the graph learning model. The GCN model training and testing are performed by NVIDIA GeForce GTX 1080 graphics card. 
%NVIDIA TITAN-XP and  

We use the leave-one-out approach for evaluation. We report test results on a circuit infected with a HT benchmark while the model is trained on other circuits and HTs. We change the test circuit and repeat training on the rest again. The process is repeated until all samples are tested. In this scenario, the circuit under test and its HT are not seen by the model in training which indicates the capability of model to locate HT in unknown circuits and HTs.
In all evaluations, we define the positive sample as Trojan node class and the negative sample as benign node class. For example, true positive represents the Trojan nodes that are correctly classified as Trojan.

%Our dataset comprises 4 base circuits that contain different HTs; RS232, AES, DES, and RC5. The RS232 is an implementation of UART serial communication transmitter and receiver. Other circuits are encryption cores with different algorithms that get an input number as plain text along with a secret key and output the encrypted number known as chipher text. The RS232 and AES samples come from the TrustHub benchmark but the RC5 and DES are open-source designs in which we inserted the Trojan circuits extracted from AES-Txx benchmarks. Our dataset embeds a variety of Trojan trigger and payload designs that leak a secret key, manipulate functionality, deteriorate the IC by draining power, or disable the chip. 

%%%%%%%%%%%%%%%%%%%%%%%%%%%%%%%%%%%%%%%%%%%%%%%%%%%%%%%%%%%%%%%%%%%%%%%%%%%%%%%%%
\begin{table}[t]
\centering
\caption{The performance of HT detection.}
\vspace{-1em}
\label{tab:detction}
\begin{tabular}{|c|c|c|c|}
\hline
\rowcolor[HTML]{AED09D} 
Circuit                    & AES  & DES  & RC5  \\ \hline
Classified as Trojan node  & 1             & 2             & 0             \\ \hline
\rowcolor[HTML]{EBF1E9} 
Classified as benign node  & 13437         & 10210         & 2106          \\ \hline
Total nodes                & 13438         & 10212         & 2106          \\ \hline
\rowcolor[HTML]{EBF1E9} 
Classified as Trojan/total & 7.44e-5       & 1.95e-4       & 0             \\ \hline
HT detection accuracy      & \multicolumn{3}{c|}{100\%}                    \\ \hline
\end{tabular}
\vspace{-1.5em}
\end{table}
%%%%%%%%%%%%%%%%%%%%%%%%%%%%%%%%%%%%%%%%%%%%%%%%%%%%%%%%%%%%%%%%%%%%%%%%%%%%%%%%%
%%%%%%%%%%%%%%%%%%%%%%%%%%%%%%%%%%%%%%%%%%%%%%%%%%%%%%%%%%%%%%%%%%%%%%%%%%%%%%%%%
\begin{table}[t]
\centering
\caption{The summary of dataset and HT localization performance.}
\vspace{-1em}
\label{tab:ave-result}
\begin{tabular}{|c|c|c|c|c|}
\hline
\rowcolor[HTML]{AED09D} 
Benchmark                   & All & AES-Txx    & DES-Txx    & RC5-Txx    \\ \hline
Accuracy                  & 99.6\%       & 99.8\% & 99.8\% & 99.2\% \\ \hline
\rowcolor[HTML]{EBF1E9} 
F1-score                  & 93.1\%       & 93.2\% & 92.2\% & 93.1\% \\ \hline
Precision                 & 99.0\%       & 99.8\% & 97.5\% & 99.4\% \\ \hline
\rowcolor[HTML]{EBF1E9} 
Recall                    & 88.0\%       & 88.0\% & 87.9\% & 88.0\% \\ \hline
\# of nodes  & 2000-14000   & 13438  & 10212  & 2106   \\ \hline
\rowcolor[HTML]{EBF1E9} 
Time & $<$ 500ms      & 222ms & 162ms & 37ms       \\ \hline
\end{tabular}
\vspace{-2em}
\end{table}
%%%%%%%%%%%%%%%%%%%%%%%%%%%%%%%%%%%%%%%%%%%%%%%%%%%%%%%%%%%%%%%%%%%%%%%%%%%%%%%%%
%%%%%%%%%%%%%%%%%%%%%%%%%%%%%%%%%%%%%%%%%%%%%%%%%%%%%%%%%%%%%%%%%%%%%%%%%%%%%%%%%
\vspace{-0.5em}
\subsection{Comparing HT Localization Methods}

In this section, we compare our model with other HT localization methods in the literature that are elaborated in Section \ref{sec:related-works}. u A quantitative comparison is challenging due to a couple of reasons. Firstly, the dataset and experiment conditions are very varied among different works. For example, \cite{hicks2010overcoming, waksman2013fanci, zhang2015veritrust, farahmandi2017trojan} papers propose various ideas to locate the HT nodes in the circuit, and they demonstrate promising results on their limited sets of benchmarks. However, each one reveals the shortcomings of the former method against distinct Trojans. Secondly, diverse techniques are used for localization with varied evaluation metrics that are not comparable. For example, \cite{islam2020framework} reports the error in activity estimation, which is further used for marking low-activity regions as vulnerable to HT. On the other hand, \cite{islam2019socio} and our approach demonstrate accuracy in finding the Trojan nodes. Although the direct quantitative comparison is not feasible, we provide numeric evaluation of different methods performance in terms of how successful they were to locate HT (using accuracy, recall, and error metrics), how many benign nodes were mislabled as HT (using false positive rate and precision metrics) in Table~\ref{tab:survey}. The comparison shows the superior perfomance of our model in sucessfully detecting HT nodes with very low false positive. The metrics definition are elaborated in Section~\ref{sec:exp-results}. 

We study them from qualitative aspects such as pre-silicon or post-silicon HT localization, golden reference-free, automated feature extraction/property definition, localization resolution, and the ability to detect various types of Trojans. According to Table~\ref{tab:survey}, our compelling model surmounts the shortcomings of the state-of-the-art. The post-silicon techniques postpone the HT localization until after fabrications, when the HT removal is very time-consuming and expensive. Therefore, it is crucial to locate and remove Trojans inserted in the design stage early before manufacturing. On the other hand, pre-silicon HT localization approaches mostly suffer from low resolutions because they cannot detect the Trojan nodes specifically. Instead, they mark the suspicious areas that are prone to HT insertion. Thus, they require further manual revision of circuit partitions to check for Trojan nodes, and their localization process is not automated.

%%%%%%%%%%%%%%%%%%%%%%%%%%%%%%%%%%%%%%%%%%%%%%%%%%%%%%%%%%%%%%%%%%%%%%%%%%%%%%%%%
%%%%%%%%%%%%%%%%%%%%%%%%%%%%%%%%%%%%%%%%%%%%%%%%%%%%%%%%%%%%%%%%%%%%%%%%%%%%%%%%%
\vspace{-1em}
\subsection{HT Detection and Localization Performance}  \label{sec:exp-results}

After finding the best model and architecture which is elaborated in Sections \ref{best-arch} and \ref{best-weight}, we construct our final model.The evaluation results per benchmark are reported in Table~\ref{tab:resultPerHW}. We also include the number of Trojan nodes and the ratio of Trojan nodes to total nodes in the graph to reflect the effect of HT size. We consider several evaluation metrics to assess the performance from different perspectives. The most common metric for classification is accuracy which expresses the correctly classified nodes over all nodes. Accuracy is intuitive but does not suffice since class distribution between nodes is not uniform. Thus, we look into the F1-score, the weighted average of recall and precision. 

Recall expresses the ability to find all Trojan nodes in a design. On the other hand, precision is an indicator of False Positive (FP) and expresses the proportion of the nodes our model labels as Trojan, actually are Trojan.The combination of precision and recall metrics examines the model's performance in detecting Trojan nodes while avoiding mislabeling benign nodes as Trojan. We count True Positive (TP), False Negative (FN), and FP and calculate these metrics as follows:

\begin{equation}
    P=\frac{TP}{TP + FP} , \text{   } R=\frac{TP}{TP + FN} 
\end{equation}
\begin{equation}
    F_{\beta}\-score=\frac{(1+\beta^2)*P*R}{\beta^2*P + R}
\end{equation}
\begin{equation}
    Accuracy =\frac{TP + TN}{TP + TN + FN + FP}
\end{equation}
\begin{equation}
    \text{HT nodes} = TP + FN
\end{equation}
\begin{equation}
    \text{HT/total ratio} = \frac{TP + FN}{TP + TN + FN + FP}
\end{equation}

We provide a summary of results in Table~\ref{tab:ave-result} in which the average of metrics are calculated for each circuit as well as average node classification time. It can be observed that high accuracy and F1-score in HT localization are maintained for all circuits regardless of size.  The computation and timing of HT localization depend on the size of the circuit. Studying the timing in diverse designs, it is observed that HT localization time scales linearly with the number of nodes in the graph representation of the circuit, which makes it scalable for large designs. In conclusion, our GCN model exhibits high performance in locating the HT nodes with low false positives (below 0.009\%) in less than 1 second.
Further, we study the performance of our model in HT detection by testing it for the HT-free circuits of AES, DES, and RC5. The number of nodes classified as Trojan/benign is mentioned in Table~\ref{tab:detction}. These results show that our model can determine if the design is healthy as it finds only few false Trojan nodes in a design graph with thousands of nodes that is negligible.
%%%%%%%%%%%%%%%%%%%%%%%%%%%%%%%%%%%%%%%%%%%%%%%%%%%%%%%%%%%%%%%%%%%%%%%%%%%%%%%%%
%%%%%%%%%%%%%%%%%%%%%%%%%%%%%%%%%%%%%%%%%%%%%%%%%%%%%%%%%%%%%%%%%%%%%%%%%%%%%%%%%
\begin{figure}[t]
\centering
\includegraphics[width=0.48\textwidth]{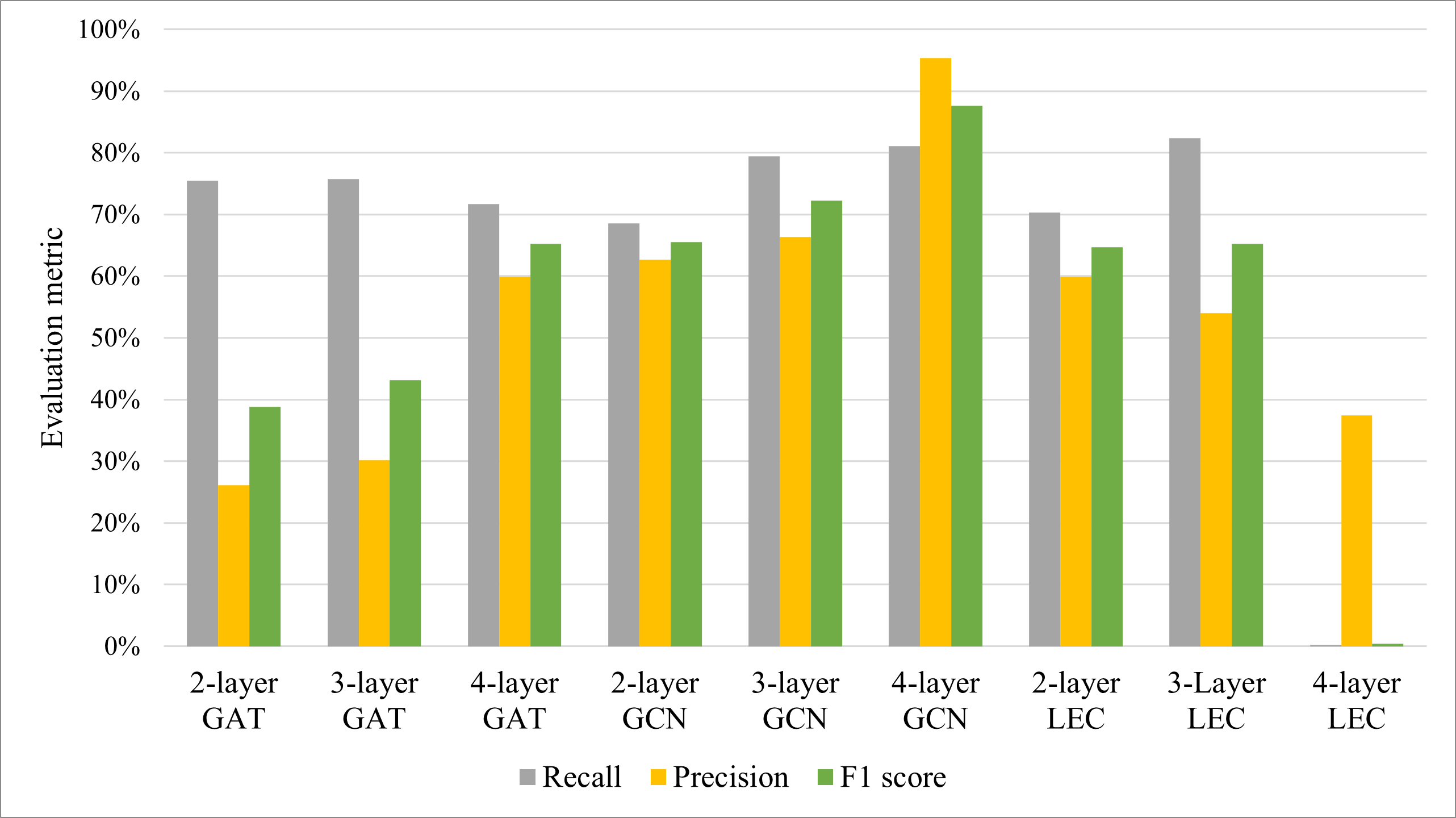}
\vspace{-1em}
\caption{Performance of various graph neural network models and architectures.}
\label{fig:models}
\vspace{-1em}
\end{figure}
%%%%%%%%%%%%%%%%%%%%%%%%%%%%%%%%%%%%%%%%%%%%%%%%%%%%%%%%%%%%%%%%%%%%%%%%%%%%%%%%%
%%%%%%%%%%%%%%%%%%%%%%%%%%%%%%%%%%%%%%%%%%%%%%%%%%%%%%%%%%%%%%%%%%%%%%%%%%%%%%%%%
\begin{table}[t]
\centering
\caption{HT localization performance, number of Trojan nodes, and their ratio to total nodes for all benchmarks.}
\vspace{-1em}
\label{tab:resultPerHW}
\begin{tabular}{|C{1.4cm}|C{0.8cm}|C{0.8cm}|C{0.8cm}|C{0.8cm}|C{0.7cm}|C{0.9cm}|}
\hline
\rowcolor[HTML]{AED09D} 
  Benchmark &
  Acc $^1$ &
  \begin{tabular}[c]{@{}c@{}}F1\\ score\end{tabular} &
  Prec $^2$ &
  Recall &
  \begin{tabular}[c]{@{}c@{}}HT\\ nodes\end{tabular} &
  \begin{tabular}[c]{@{}c@{}}HT/total\\ ratio\end{tabular} \\ \hline
AES-T100  & 100\% & 99.4\% & 100\% & 98.8\% & 481 & 3.46\%  \\ \hline
\rowcolor[HTML]{EBF1E9} 
AES-T200  & 100\% & 99.4\% & 100\% & 98.8\% & 486 & 3.40\%  \\ \hline
AES-T300  & 99.5\%  & 94.7\% & 98.6\%  & 91.0\% & 635 & 5.33\%  \\ \hline
\rowcolor[HTML]{EBF1E9} 
AES-T400  & 99.9\%  & 94.2\% & 100\% & 89.1\% & 110 & 0.71\%  \\ \hline
AES-T500  & 99.8\%  & 85.4\% & 100\% & 74.5\% & 94  & 0.69\%  \\ \hline
\rowcolor[HTML]{EBF1E9} 
AES-T600  & 99.9\%  & 89.9\% & 100\% & 81.6\% & 87  & 0.64\%  \\ \hline
AES-T700  & 99.8\%  & 97.8\% & 100\% & 95.7\% & 562 & 3.98\%  \\ \hline
\rowcolor[HTML]{EBF1E9} 
AES-T800  & 99.8\%  & 97.6\% & 100\% & 95.2\% & 628 & 4.42\%  \\ \hline
AES-T900  & 99.8\%  & 97.8\% & 99.8\%  & 96.0\% & 569 & 4.03\%  \\ \hline
\rowcolor[HTML]{EBF1E9} 
AES-T1000 & 99.9\%  & 98.4\% & 100\% & 96.8\% & 503 & 3.73\%  \\ \hline
AES-T1100 & 99.9\%  & 97.7\% & 100\% & 95.6\% & 568 & 3.21\%  \\ \hline
\rowcolor[HTML]{EBF1E9} 
AES-T1200 & 99.9\%  & 98.3\% & 100\% & 96.7\% & 509 & 3.51\%  \\ \hline
AES-T1300 & 99.1\%  & 87.5\% & 100\% & 77.8\% & 688 & 3.87\%  \\ \hline
\rowcolor[HTML]{EBF1E9} 
AES-T1400 & 99.5\%  & 94.1\% & 98.8\%  & 89.9\% & 723 & 4.05\%  \\ \hline
AES-T1500 & 99.5\%  & 92.1\% & 98.6\%  & 86.4\% & 664 & 3.15\%  \\ \hline
\rowcolor[HTML]{EBF1E9} 
AES-T1600 & 99.9\%  & 92.2\% & 100\% & 85.5\% & 179 & 0.82\%  \\ \hline
AES-T1700 & 99.9\%  & 89.0\% & 100\% & 80.2\% & 86  & 0.45\%  \\ \hline
\rowcolor[HTML]{EBF1E9} 
AES-T1800 & 99.9\%  & 83.3\% & 100\% & 71.4\% & 27  & 0.17\%  \\ \hline
AES-T1900 & 100\% & 82.8\% & 100\% & 70.6\% & 34  & 0.17\%  \\ \hline
DES-T100  & 99.9\%  & 99.2\% & 99.8\%  & 98.5\% & 481 & 4.50\%  \\ \hline
\rowcolor[HTML]{EBF1E9} 
DES-T200  & 99.9\%  & 99.2\% & 99.8\%  & 98.6\% & 486 & 4.54\%  \\ \hline
DES-T400  & 99.9\%  & 94.3\% & 98.0\%  & 90.9\% & 110 & 1.07\%  \\ \hline
\rowcolor[HTML]{EBF1E9} 
DES-T500  & 99.8\%  & 84.8\% & 98.6\%  & 74.5\% & 94  & 0.91\%  \\ \hline
DES-T600  & 99.8\%  & 90.0\% & 98.6\%  & 82.8\% & 87  & 0.84\%  \\ \hline
\rowcolor[HTML]{EBF1E9} 
DES-T700  & 99.8\%  & 97.6\% & 99.8\%  & 95.6\% & 562 & 5.22\%  \\ \hline
DES-T800  & 99.6\%  & 96.7\% & 98.2\%  & 95.2\% & 628 & 5.79\%  \\ \hline
\rowcolor[HTML]{EBF1E9} 
DES-T900  & 99.7\%  & 97.4\% & 99.6\%  & 95.3\% & 569 & 5.28\%  \\ \hline
DES-T1000 & 99.8\%  & 98.1\% & 99.6\%  & 96.6\% & 503 & 4.69\%  \\ \hline
\rowcolor[HTML]{EBF1E9} 
DES-T1100 & 99.7\%  & 96.9\% & 98.2\%  & 95.6\% & 568 & 5.27\%  \\ \hline
DES-T1200 & 99.8\%  & 98.2\% & 99.6\%  & 96.9\% & 509 & 4.75\%  \\ \hline
\rowcolor[HTML]{EBF1E9} 
DES-T1600 & 99.7\%  & 91.6\% & 98.7\%  & 85.5\% & 179 & 1.72\%  \\ \hline
DES-T1700 & 99.7\%  & 83.6\% & 87.3\%  & 80.2\% & 86  & 0.84\%  \\ \hline
\rowcolor[HTML]{EBF1E9} 
DES-T1800 & 99.9\%  & 80.2\% & 90.0\%  & 72.3\% & 27  & 0.26\%  \\ \hline
DES-T1900 & 99.9\%  & 81.4\% & 96.0\%  & 70.6\% & 34  & 0.33\%  \\ \hline
RC5-T100  & 99.8\%  & 99.4\% & 100\% & 98.8\% & 481 & 18.59\% \\ \hline
\rowcolor[HTML]{EBF1E9} 
RC5-T200  & 99.8\%  & 99.4\% & 100\% & 98.8\% & 486 & 18.76\% \\ \hline
RC5-T400  & 99.5\%  & 94.2\% & 100\% & 89.1\% & 110 & 4.96\%  \\ \hline
\rowcolor[HTML]{EBF1E9} 
RC5-T500  & 98.9\%  & 85.4\% & 100\% & 74.5\% & 94  & 4.27\%  \\ \hline
RC5-T600  & 99.3\%  & 90.0\% & 98.6\%  & 82.8\% & 87  & 3.97\%  \\ \hline
\rowcolor[HTML]{EBF1E9} 
RC5-T700  & 99.1\%  & 97.9\% & 100\% & 95.9\% & 562 & 21.06\% \\ \hline
RC5-T800  & 98.9\%  & 97.5\% & 99.7\%  & 95.4\% & 628 & 22.96\% \\ \hline
\rowcolor[HTML]{EBF1E9} 
RC5-T900  & 99.0\%  & 97.5\% & 99.6\%  & 95.4\% & 569 & 21.27\% \\ \hline
RC5-T1000 & 99.3\%  & 98.2\% & 100\% & 96.4\% & 503 & 19.28\% \\ \hline
\rowcolor[HTML]{EBF1E9} 
RC5-T1100 & 99.1\%  & 97.7\% & 100\% & 95.6\% & 568 & 21.24\% \\ \hline
RC5-T1200 & 99.3\%  & 98.1\% & 99.8\%  & 96.5\% & 509 & 19.46\% \\ \hline
\rowcolor[HTML]{EBF1E9} 
RC5-T1600 & 98.8\%  & 91.9\% & 99.4\%  & 85.5\% & 179 & 7.83\%  \\ \hline
RC5-T1700 & 99.2\%  & 88.5\% & 98.6\%  & 80.2\% & 86  & 3.93\%  \\ \hline
\rowcolor[HTML]{EBF1E9} 
RC5-T1800 & 99.6\%  & 83.3\% & 95.2\%  & 74.1\% & 27  & 1.27\%  \\ \hline
RC5-T1900 & 99.5\%  & 82.8\% & 100\% & 70.6\% & 34  & 1.59\%  \\ \hline
\rowcolor[HTML]{EBF1E9} 
 Average & 99.6\%  & 93.1\% & 98.9\%  & 88.4\% & 356  & 5.84\%  \\ \hline
\end{tabular}
\footnotesize{}{$^1$Accuracy  $^2$Precision}
\vspace{-2em}
\end{table}
%%%%%%%%%%%%%%%%%%%%%%%%%%%%%%%%%%%%%%%%%%%%%%%%%%%%%%%%%%%%%%%%%%%%%%%%%%%%%%%%%
%%%%%%%%%%%%%%%%%%%%%%%%%%%%%%%%%%%%%%%%%%%%%%%%%%%%%%%%%%%%%%%%%%%%%%%%%%%%%%%%%
\begin{figure}[t]
\centering
\includegraphics[width=0.48\textwidth]{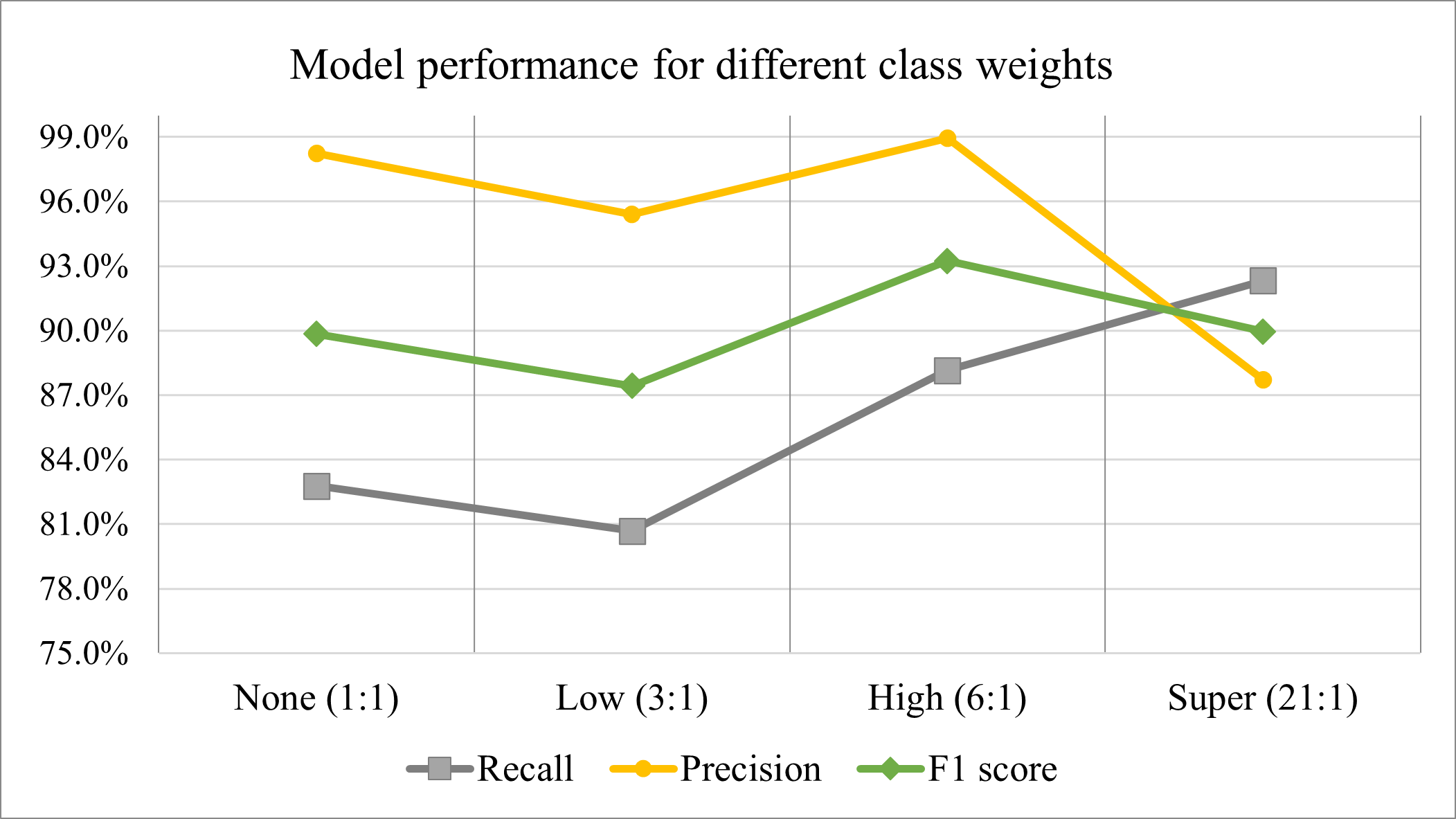}
\vspace{-1em}
\caption{Performance of GCN model with different class weights. }
\label{fig:weight}
\vspace{-1.5em}
\end{figure}
%%%%%%%%%%%%%%%%%%%%%%%%%%%%%%%%%%%%%%%%%%%%%%%%%%%%%%%%%%%%%%%%%%%%%%%%%%%%%%%%%
\vspace{-0.5em}
\subsection {The Best Graph Neural Network Architecture} \label{best-arch}

There are plenty of graph neural networks candidates with various hyper-parameters to choose as our node classification model. Thus, we devise an experiment in which we construct and test different models to find the best model and architecture for our application. In this experiment, we implement 3 different graph learning model including GCN \cite{kipf2016semi}, graph attention network (GAT) \cite{velivckovic2017graph}, and local extrema convolution (LEC) \cite{ranjan2020asap} with different architectures (2-layer to 4-layer). The evaluation results are illustrated in Figure~\ref{fig:models}. F1-score is the main evaluation metrics for comparison because it is the average of precision and recall and represents the two key expected qualities; detecting all Trojan nodes and having low false positives. The LEC model shows the worst performance, and by increasing the number of layers, its performance drops. On the contrary, the GCN and GAT models are improved by stacking more layers while GCN relatively exhibits better performance. Therefore, The GCN model with four layers is chosen for node classification.

%%%%%%%%%%%%%%%%%%%%%%%%%%%%%%%%%%%%%%%%%%%%%%%%%%%%%%%%%%%%%%%%%%%%%%%%%%%%%%%%%
\vspace{-0.5em}
\subsection {Compensation for Unbalanced Dataset} \label{best-weight}

A standard step of developing machine learning models is to find the best settings for the model based on the problem. One of the challenges of Trojan localization is the small size of the HT circuit that results in an unbalanced dataset for machine learning. In our dataset, the ratio of HT nodes to total nodes is between 0.001-0.020 (refer to Table~\ref{tab:resultPerHW}), which means the distribution of node classes is not uniform, and one class of nodes is more common. The unbalanced dataset can affect the model's performance and push it to label all nodes as the dominant class, the benign node class. To tackle this problem, we assign a higher weight to the Trojan class in loss calculation that compensates for the minority of Trojan nodes and forces the model to label more nodes as Trojan. We devise an experiment to find the optimum value for class weight by altering the relative weight of Trojan class to benign class among these values: 1:1 (none), 3:1 (low), 6:1 (high), and 21:1 (super). In the evaluation results in Figure~\ref{fig:weight}, we notice that increasing the weight of Trojan continuously increases the recall as more Trojan nodes are found. Still, after some point, it deteriorates the overall performance (F1-score) as the false positive sample increases, and consequently, the precision drops. Therefore, The best class weight with the highest F1-score is 6:1 and we use this value for further evaluations.
%%%%%%%%%%%%%%%%%%%%%%%%%%%%%%%%%%%%%%%%%%%%%%%%%%%%%%%%%%%%%%%%%%%%%%%%%%%%%%%%%
\vspace{-1em}
\section{Conclusion}

In this paper, we create a novel, golden reference-free HT localization methodology that converts the hardware design to a graph,  performs node classification on it using GCN, and outputs the malicious circuit corresponding to Trojan nodes. Our methodology is fully automated without any need for manual feature extraction or code inspection. Our evaluation demonstrate that it locate Trojan with 99.6\% accuracy, 93.1\% F1-score, and false positive rate below 0.009\%.
\vspace{-1em}
% \cleardoublepage
\bibliographystyle{IEEEtran}
\bibliography{bibliography}

\end{document}